\documentclass[pdflatex,sn-mathphys-num,authoryear]{sn-jnl}

\usepackage{graphicx}%
\usepackage{multirow}%
\usepackage{amsmath,amssymb,amsfonts}%
\usepackage{amsthm}%
\usepackage{mathrsfs}%
\usepackage[title]{appendix}%
\usepackage{xcolor}%
\usepackage{textcomp}%
\usepackage{manyfoot}%
\usepackage{booktabs}%
\usepackage{algorithm}%
\usepackage{algorithmicx}%
\usepackage{algpseudocode}%
\usepackage{listings}%
\usepackage{xurl}  

\theoremstyle{thmstyleone}%
\theoremstyle{thmstyletwo}%

\theoremstyle{thmstylethree}%

\usepackage[table]{xcolor}
\usepackage{booktabs}
\usepackage{array}
\usepackage{makecell}
\usepackage{caption}
\usepackage{pgfplots}
\usepackage{tikz}

\pgfplotsset{compat=newest}

\newcommand{\colorFermentation}[1]{%
  \ifdim #1 pt < 60 pt \cellcolor[RGB]{252,244,240}
  \else\ifdim #1 pt < 80 pt \cellcolor[RGB]{240,220,200}
  \else\ifdim #1 pt < 90 pt \cellcolor[RGB]{220,190,160}
  \else \cellcolor[RGB]{200,150,100}
  \fi\fi\fi #1}
  
\newcommand{\colorMoisture}[1]{%
  \ifdim #1 pt < 4.7 pt \cellcolor[RGB]{255,255,240}
  \else\ifdim #1 pt < 5.0 pt \cellcolor[RGB]{255,245,180}
  \else\ifdim #1 pt < 5.3 pt \cellcolor[RGB]{255,235,130}
  \else \cellcolor[RGB]{255,220,90}
  \fi\fi\fi #1}

\newcommand{\colorCadmium}[1]{%
  \ifdim #1 pt < 1.5 pt \cellcolor[RGB]{255,252,252}
  \else\ifdim #1 pt < 2.5 pt \cellcolor[RGB]{255,220,220}
  \else\ifdim #1 pt < 3.5 pt \cellcolor[RGB]{255,170,170}
  \else \cellcolor[RGB]{245,120,120}
  \fi\fi\fi #1}

\newcommand{\colorPolyphenols}[1]{%
  \ifdim #1 pt < 28 pt \cellcolor[RGB]{245,255,245}
  \else\ifdim #1 pt < 35 pt \cellcolor[RGB]{210,245,210}
  \else\ifdim #1 pt < 38 pt \cellcolor[RGB]{170,230,170}
  \else \cellcolor[RGB]{100,200,100}
  \fi\fi\fi #1}

\newcommand{\colorHours}[1]{%
  \ifdim #1 pt < 50 pt \cellcolor[RGB]{245,255,255}
  \else\ifdim #1 pt < 100 pt \cellcolor[RGB]{215,245,245}
  \else\ifdim #1 pt < 200 pt \cellcolor[RGB]{180,230,230}
  \else \cellcolor[RGB]{110,200,200}
  \fi\fi\fi #1}

\renewcommand{\arraystretch}{1.25}
\newcommand{\bacca}[1]{{\textcolor{black}{#1}}}
\newcommand{\bemc}[1]{{\textcolor{black}{#1}}}
\definecolor{yelo}{RGB}{255, 100, 0} 
\newcommand{\kebin}[1]{\textcolor{black}{#1}}
\newcommand{\emma}[1]{{\textcolor{black}{#1}}}

\begin{document}

\title[Article Title]{Learning-based Spectral Regression for Cocoa Bean Physicochemical Property Prediction}


\author[1]{\fnm{Kebin} \sur{Contreras} }
\equalcont{These authors contributed equally to this work.}

\author[2]{\fnm{Emmanuel} \sur{Martinez} }
\equalcont{These authors contributed equally to this work.}

\author[2]{\fnm{Brayan} \sur{Monroy} }

\author[2]{\fnm{Sebastian} \sur{Ardila}}

\author[2]{\fnm{Cristian} \sur{Ramirez}}

\author[2]{\fnm{Mariana} \sur{Caicedo}}

\author[1]{\fnm{Hans} \sur{Garcia} }

\author[3]{\fnm{Tatiana} \sur{Gelvez-Barrera} }

\author[2]{\fnm{Juan} \sur{Poveda-Jaramillo} }

\author[2]{\fnm{Henry} \sur{Arguello} }

\author*[2]{\fnm{Jorge} \sur{Bacca}}\email{jbacquin@uis.edu.co}


\affil[]{%
  \orgdiv{$^{1}$Physics School and $^{2}$Department of Computer Science}, 
  \orgname{Universidad Industrial de Santander}, 
  \orgaddress{\street{Carrera 27 Calle 9}, 
              \city{Bucaramanga}, 
              \postcode{680001}, 
              \state{Santander}, 
              \country{Colombia}}}

\affil[3]{\orgdiv{CNRS, Inserm, CREATIS UMR 5220}, \orgname{Université de Lyon, Université Claude Bernard Lyon 1, UJM-Saint Etienne}, \orgaddress{\street{Street}, \city{Lyon}, \postcode{69000}, \state{ Auvergne-Rhône-Alpes}, \country{France}}}


\abstract{Cocoa bean quality assessment is essential for ensuring compliance with commercial standards, protecting consumer health, and increasing the market value of the cocoa product. The quality assessment estimates key physicochemical properties, such as fermentation level, moisture content, polyphenol concentration, and cadmium content, among others. This assessment has traditionally relied on the accurate estimation of these properties via visual or sensory evaluation, jointly with laboratory-based physicochemical analyses, which are often time-consuming, destructive, and difficult to scale. This creates the need for rapid, reliable, and noninvasive alternatives. Spectroscopy, particularly in the visible (VIS: 400–700 nm) and near-infrared (NIR: 700–2500 nm) ranges, offers a non-invasive alternative by capturing the molecular signatures associated with these properties. Therefore, this work introduces a scalable methodology for evaluating the quality of cocoa beans by predicting key physicochemical properties from the spectral signatures of cocoa beans. This approach utilizes a conveyor belt system integrated with a VIS-NIR spectrometer, coupled with learning-based regression models. Furthermore, a dataset is built using cocoa bean batches from Santander, Colombia. Ground-truth reference values were obtained through standardized laboratory analyses and following commercial cocoa quality regulations. To further evaluate the proposed methodology's generalization, performance is tested on samples collected from other Colombian regions and from Cusco, Peru. Experimental results show that the proposed models achieved $\mathcal{R}^2$ scores exceeding $0.98$ across all physicochemical properties, and reached $0.96$ accuracy on geographically independent samples. This non-destructive approach represents a suitable and scalable alternative to conventional laboratory methods for quality assessment across the cocoa production chain.}


\maketitle

\section*{Introduction}\label{Introduction}

Cocoa beans \emma{ (\textit{Theobroma cacao} L.)} rank among the world's leading agricultural products, with global production estimated at 4.368 million tons during the 2023/2024 crop season \emma{according to the International Cocoa Organization}~\citep{icco2025may}. Cultivation \emma{remains} concentrated in West Africa ($\approx75\%$),  Asia and Oceania ($\approx5\%$), and Latin America ($20\%$) \citep{kongor2024cocoa}. \emma{In Latin America and the Caribbean, cocoa occupies over 1.7 million hectares and contributes close to one-fifth of global supply, with major producing countries including Ecuador, Brazil, Peru, Colombia, the Dominican Republic, and Mexico\citep{icco2025may, huetz2022cocoa}. The region is also the world’s leading source of fine-flavor cocoa, accounting for approximately 90\% of global exports in this premium segment, particularly through Ecuador, the Dominican Republic, and Peru \citep{icco2023fine}.}

\emma{Critically, cocoa farming supports the livelihoods of approximately 5 to 6 million smallholder households worldwide, often operating low-input farms of 2 to 5 hectares, and contributes 60\% to 90\% of their household income, boosting food security, education, and rural economies~\citep{kongor2024cocoa, huetz2022cocoa}. For these farmers, participation in fine-flavor cocoa markets offers access to higher premiums at origin, although the small size of this niche market (only 12\% of global exports) and strict quality requirements present barriers \citep{icco2023fine}. Nonetheless, cocoa farming remains a culturally rooted and economically vital activity, particularly in Latin America, offering smallholder farmers a path to resilience, rural development, and greater inclusion in sustainable and differentiated value chains.}

The quality of cocoa beans is determined by genetic factors, post-harvest practices, and geographic conditions, among others, which play a crucial role in shaping their final attributes~\citep{kongor2016factors}. Key post-harvest processes, such as, fermentation, drying, and roasting, directly influence the sensory profile and physicochemical characteristics of the beans \citep{de2001structural}. These processes affect critical attributes like flavor, aroma, and nutritional content, which in turn determine the product's marketability and economic value~\citep{de2016cocoa}. Therefore, the ability to assess and \bemc{determine} these quality factors is essential for maintaining consistency in production and meeting the standards and requirements of both national and international markets.

\emma{In Latin America, countries such as Colombia, Ecuador, Brazil, and Peru have established national standards to regulate cocoa quality by international trade and food safety requirements. These regulations define criteria such as fermentation level, moisture content, cadmium concentration, and physical defects. Colombia applies the NTC 1252:2021, which classifies dry cocoa beans by quality \citep{ntc1252_2021}. Ecuador enforces NTE INEN 176:2021, its official standard for fermented cocoa beans \citep{inen176_2021}. Brazil regulates quality through MAPA Normative No. 38/2018, which sets classification parameters for cocoa beans \citep{mapa38_2018}. In Peru, cocoa quality control relies on NTP-ISO 2292:2019, based on international sampling guidelines \citep{ntp2292_2019}.}

\begin{figure}[!t]
    \centering
    \includegraphics[width=0.5\linewidth]{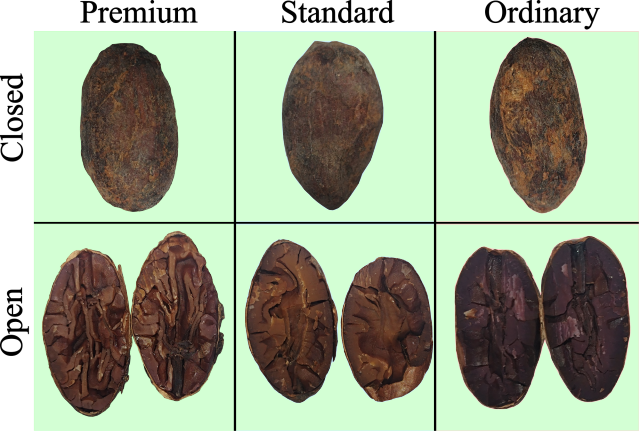}
    \caption{\kebin{Visual classification of cocoa beans based on the cut test method. The top row shows closed beans \bacca{being difficult to classify}, and the bottom row shows the corresponding internal appearance after cutting. Beans are categorized as Premium, Standard, or Ordinary based on internal and external appearance, considering color, texture, and uniformity.}}
    \label{fig:cocoa_beans}
\end{figure}

\emma{Fermentation remains a key determinant of cocoa quality across the region. It is traditionally evaluated through the cut test, a destructive visual inspection method rooted in ancestral knowledge \citep{ntc1252_2021, inen176_2021, mapa38_2018, ntp2292_2019}. Although widely used, this technique is inherently subjective, relying on human interpretation of internal and external bean characteristics. As illustrated in Fig.~\ref{fig:cocoa_beans}, cocoa beans are visually classified into premium, standard, or ordinary categories based on external appearance and internal characteristics revealed by the cut test. In particular, ordinary beans display a dark purple interior, indicative of low fermentation. In contrast, standard and premium beans show more developed brown coloration and well-formed internal structures, reflecting higher levels of fermentation, with premium beans typically exhibiting the most uniform and desirable traits \citep{ntc1252_2021, inen176_2021, mapa38_2018, ntp2292_2019}.}

In addition to visual inspection, physicochemical properties such as moisture content, polyphenol concentration, and cadmium levels are important indicators of cocoa bean quality \citep{samanta2022dark, abt2020perspective}. Moisture affects shelf life and the risk of mold growth, while polyphenols are valued for their antioxidant properties but are often degraded during fermentation and roasting \citep{samanta2022dark}. Cadmium, which is influenced by soil composition, presents a food safety concern, particularly in beans from some Latin American regions where soil levels are naturally higher \citep{abt2020perspective}. These factors impact both the nutritional and sensory qualities of cocoa and have significant implications for regulatory standards and international trade \citep{samanta2022dark, abt2020perspective}.

Cocoa quality assessment typically relies on laboratory-based quantitative techniques such as atomic absorption spectroscopy (AAS)~\citep{araujo2020verification}, gas chromatography (GC)~\citep{ducki2008evaluation}, and mass spectrometry (MS)~\citep{cain2019food}. While these methods provide detailed insights into the internal composition and physicochemical properties of cocoa beans, they are invasive and destructive. Additionally, they require specialized infrastructure, trained personnel, and incur high operational costs, making them largely inaccessible to rural communities and small-scale agriculture~\citep{niemenak2014physical}. The extended processing times also limit their applicability in scenarios where rapid decision-making is needed. Consequently, many small-scale producers are unable to access such analyses, hindering their ability to optimize cocoa quality and remain competitive in the market.

Recent research has shown that visible spectrum (VIS) \bacca{ranged from 400-700 nm} and near-infrared (NIR) \bacca{range from 700-2500 nm} spectroscopy are effective, reliable, and non-invasive techniques for \bacca{predicting the cocoa quality} \citep{sanchez2021classification,hashimoto2018quality, araujo2020verification,suarez2025automated,pinto2024advances, diaz2025deep,teye2020cocoa}.
These methods collect data reflecting complex molecular vibrations that reveal the physicochemical properties of cocoa beans~\citep{sandorfy2007principles}. In the VIS range, the reflected radiation from the material is observed, while the absorbed radiation induces electronic transitions in the valence electrons of the constituent molecules. In the NIR range, the overtones of molecular bond vibrations become detectable. In particular, spectral information has been used to predict quality-related attributes such as fermentation level, polyphenol content, and antioxidant activity, with strong correlations reported between polyphenol levels and fermentation \citep{sanchez2021classification, gomez2019non, caporaso2018hyperspectral, alvarado2023emerging, DiazDelgado2025}. Most recent studies on cocoa quality assessment have been conducted in African countries, where regional environmental conditions and cocoa varieties significantly influence the analysis and outcomes\bacca{, often requiring the destruction of the beans, since conventional laboratory protocols depend on grinding or chemically processing the samples to obtain accurate measurements}~\citep{ferraris2023machine, ashiagbor2020pixel, musah2019assessment, hashimoto2018quality}. 

\emma{Therefore, this study presents a comprehensive framework for estimating the quality of Colombian cocoa beans using noninvasive spectral analysis. It introduces a standardized cocoa bean acquisition protocol and a custom-designed spectral optical system that predicts key physicochemical properties such as fermentation level, moisture content, cadmium concentration, and polyphenol concentration directly from spectral data. Unlike traditional methods that rely on destructive sampling or laboratory processing, this approach preserves the physical integrity of the beans and is suitable for post-harvest analysis. A spectral dataset was collected from cocoa beans in Santander, Colombia, covering wavelengths from 400 to 2500 nanometers with a spectral resolution of 3648 digitized points. Ground truth labels were obtained through standardized laboratory analyses.}

\emma{To benchmark predictive performance, a comparative framework was developed to evaluate state-of-the-art machine learning and deep learning regression models in a supervised setting. This enables accurate, scalable, and noninvasive quality assessment for the Colombian cocoa industry. To assess model generalization, additional samples were collected from other regions of Colombia such as Putumayo, Huila and Santander, as well as from Cusco in Peru. The experimental evaluation demonstrated that the proposed methods consistently delivered highly reliable predictions across all physicochemical properties, maintaining strong performance even when tested on geographically distinct samples. This non-destructive strategy therefore provides a practical and scalable alternative to conventional laboratory techniques for quality assessment throughout the cocoa production chain.}

\section*{Materials and Methods}\label{sec11}

\subsection*{Physicochemical Cocoa Bean Labeling Settings}

To assess cocoa bean quality, this study considers key physicochemical attributes: fermentation level, moisture content, polyphenol concentration, and cadmium content. These properties are critical indicators of both product quality and compliance with safety standards. Each batch of cocoa beans was analyzed following the proposed protocol, and labeled accordingly based on these measurements. Fermentation level was evaluated visually through cut tests as established in national standards. Moisture, polyphenol content, and cadmium concentration were quantified through laboratory methods conducted by the Grupo de Ciencia y Tecnología de Alimentos (CICTA)~\citep{CICTA_UIS_GrupLAC_2025} at the UIS. These analyses follow validated procedures and relevant technical standards, including national (NTC) and international norms. Table~\ref{tab:physicochemical_final} summarizes the relevance of each property and the analytical methods employed for their determination.

\begin{table*}[t]
\centering
\caption{Physicochemical properties, analytical methods and regulations for cocoa bean quality assessment.}
\label{tab:physicochemical_final}
\begin{tabular}{
>{\centering\arraybackslash}p{2.2cm}
>{\raggedright\arraybackslash}p{7cm}
>{\centering\arraybackslash}p{3cm}
}
\toprule \toprule
\textbf{Property} & \textbf{Importance} & \textbf{Method} / \textbf{Regulation} \\
\midrule

\textbf{Fermentation Level} & Influences flavor, aroma, and bean color. Evaluated externally (uniformity, mold) and internally via “cut test.” Reddish-brown beans indicate proper fermentation and purple implies incomplete processing. & Cut test (visual) \textit{NTC 1252:2021} \citep{ntc1252_2021} \\

\midrule

\textbf{Moisture} & Reflects drying efficiency and influences shelf life. Elevated moisture content increases the risk of microbial growth, particularly molds. Determined by gravimetric analysis through weight loss after oven-drying at 103±2°C. & Gravimetric method. \textit{GOMESL.01 V07, 2023-06-26}~\citep{CICTA_UIS_GrupLAC_2025}; \textit{NTC 1252:2021}~\citep{ntc1252_2021} \\

\midrule

\textbf{Polyphenols} & Key indicators of antioxidant capacity and contributors to sensory attributes. Quantified via their radical-scavenging activity against DPPH and ABTS, monitored by the decrease in absorbance. & UV-Vis spectrophotometry \citep{kus1996derivative}. \textit{GOMEPT.01 V01, 2021-09-23}~\citep{CICTA_UIS_GrupLAC_2025} \\

\midrule

\textbf{Cadmium Content} & Toxic heavy metal subject to strict regulatory limits due to its bioaccumulative nature. Typically absorbed from contaminated soils and evaluated using microwave-assisted digestion followed by atomic absorption spectroscopy (AAS). &  \textit{NTC-EN 14084:2021}~\citep{icontec2021ntc_en14084}. \\

\bottomrule \bottomrule
\end{tabular}
\end{table*}

\subsection*{Spectral Acquisition System}

The system used to acquire spectral signatures of cocoa beans consists of a halogen light source (HL-200) with a spectral coverage of 340–2400 nm, connected to a bifurcated optical fiber (R200-7-VIS-NIR). One arm of the bifurcated fiber redirects the concentrated light beam from the halogen source to the cocoa bean, while the other arm collects the reflected light to be dispersed and integrated by linear sensors inside the Vis-NIR (FLAME-S-VIS-NIR-ES) and NIR-SWIR (NIRQUEST+2.5) spectrometers. The beans are transported through the optical path by a custom-designed conveyor belt powered by a NEMA17 stepper motor and controlled by an Arduino UNO using an A4988 stepper motor driver. The bifurcated fiber is fixed 14 cm (measured from its output edge) above the conveyor belt surface. A 3D-printed funnel and pinion system ensures proper alignment and movement of the cocoa beans throughout the spectral acquisition process. The system captures at least 30 spectral signatures per bean as it moves at a speed of 45.3 mm/s.

\subsection*{Dataset Construction}

The dataset contains VIS and NIR spectral signatures along with their corresponding physicochemical labels. Reflectance values were calculated relative to white and black references obtained from a Spectralon. To eliminate noisy boundary regions, the spectral data were cropped to the (500–800 nm) range for VIS and the (1100–2000 nm) range for NIR.

Spectral signatures unrelated to cocoa beans (e.g., background regions) were discarded using a threshold of 0.25 based on the SAM method~\citep{kruse1993spectral}. The physicochemical labels correspond to a subset of spectral signatures obtained from the same physical batch of cocoa beans.

To reduce variance and improve robustness, a bootstrapping \citep{zoubir2007bootstrap} resampling procedure was applied to each cocoa batch. This involved sampling $K$ random subsets of size $s$ from the available signatures and computing the mean for each subset. Specifically, for VIS data, 1000 realizations were generated from an initial set of 1000 signatures, and for NIR, 2000 realizations were produced from 500 original signatures. In both cases, the sample size was fixed at $s = 50$.

\subsection*{Model Training and Evaluation}

All experiments were conducted on a workstation equipped with an AMD Ryzen 7 5700X 8-core processor operating at 3.40 GHz, 64 GB of RAM, and an NVIDIA GeForce RTX 4070 GPU with 12 GB of VRAM.

The evaluated regression models included Support Vector Regression (SVR)\citep{smola2004tutorial}, Random Forest\citep{breiman2001random}, CNN \citep{liu2019spectrum}, Long Short-Term Memory (LSTM)\citep{hochreiter1997long} networks, SpectralNet \citep{martins2022spectranet}, and Transformer-based architectures\citep{vaswani2017attention}. Models were trained separately using input features derived from either the visible (500–800 nm) or near-infrared (1100–2000 nm) spectral ranges. Evaluation metrics comprised the coefficient of determination ($\mathcal{R}^2$) and Mean Squared Error ($MSE$), calculated for each physicochemical property.

Hyperparameter optimization was performed using grid search and cross-validation on the training set. Deep learning models were implemented in PyTorch and trained using the Adam optimizer, with early stopping criteria based on validation loss convergence.

\section*{Results and Discussion}\label{sec2}

\subsection*{Cocoa Bean Acquisition Protocol}
\label{subsec:protocol}

\begin{figure}[h]
    \centering
    \includegraphics[width=1\linewidth]{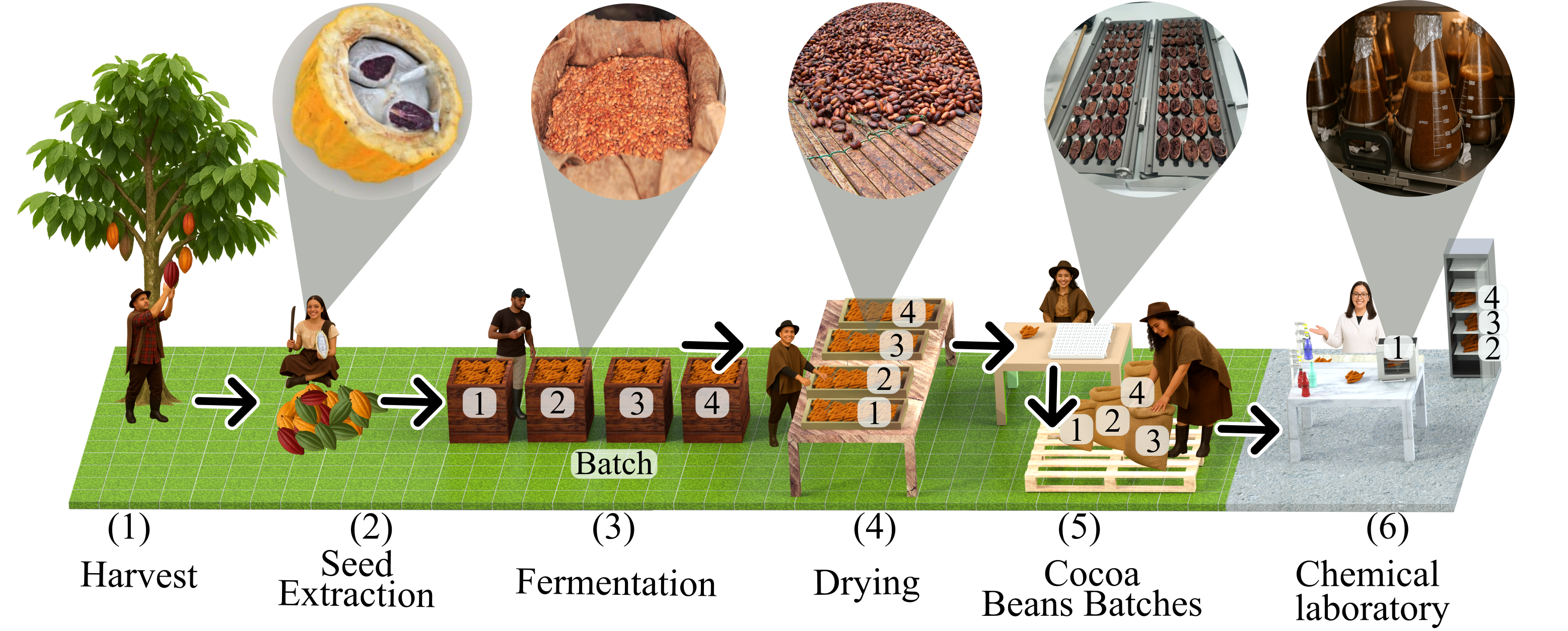}
    \caption{End-to-end pipeline for cocoa bean acquisition and labeling: (1) Harvest of ripe pods; (2) Manual seed extraction; (3) Controlled fermentation in wooden boxes with banana leaves; (4) Sun-drying to stable weight; (5) Guillotine cut test for visual fermentation grading and packaging; (6) Batch split for laboratory analyses.}
    \label{fig:cocoa-acquisition}
\end{figure}

\kebin{The acquisition and labeling protocol of cocoa beans was structured into seven sequential stages, as illustrated in Fig.~\ref{fig:cocoa-acquisition}. This workflow was designed to ensure a reproducible and standardised pipeline from field collection to chemical and spectral laboratory analysis. A batch is defined as the experimental unit, consisting of 1.5~kg of dried cocoa beans. \emma{All procedures were carried out independently for each batch as described below:}}

\kebin{\textbf{Stage 1 – Harvest:} Cocoa pods were manually harvested at optimal ripeness from a farm located in El Carmen de Chucurí, Santander, Colombia (coordinates $6^{\circ}41^{\prime}53^{\prime\prime}$N, $73^{\circ}30^{\prime}40^{\prime\prime}$W). The farm is part of \textit{“La Asociación de Campesinos Vecinos del Parque Natural Nacional Serranía Los Yariguíes”} (ASOCAPAYARI), and the beans were supplied by a farmer affiliated with the \emma{Federación Nacional de Cacaoteros (FEDECACAO) in Colombia}. All pods mainly correspond to three commercial clones widely cultivated in Colombia: CNN-51, ICS-95, and TCS-01 \citep{rosasresponse, rodriguez2019cacao}.}

\kebin{\textbf{Stage 2 – Seed Extraction:} The harvested pods were opened, and cocoa beans were manually extracted by trained personnel to maintain consistency across batches.}

\kebin{\textbf{Stage 3 – Fermentation:} The extracted beans were placed in wooden boxes lined with banana leaves and fermented \emma{under controlled temperature conditions, following regional fermentation practices that incorporate natural microbial activity and periodic mixing to ensure uniformity}. To ensure variation in fermentation levels across batches, four different durations were used: 96, 144, 192, and 264 hours. This variation was critical for assessing the influence of fermentation on physicochemical properties.}

\kebin{\textbf{Stage 4 – Drying:} After fermentation, the beans were \emma{sun-}dried until reaching a constant weight. \emma{Drying was performed} under controlled conditions in accordance with Colombian regulation NTC~1252:2021, which stipulates a maximum moisture content of 10\% for dried cocoa beans.}

\kebin{\textbf{Stage 5 – Guillotine Cut Test:} To \emma{assess} fermentation level, \emma{each batch} were evaluated using the NTC~1252:2021 standard. A Swiss guillotine was used to bisect 100 beans per batch. The beans were classified into three categories: premium (P, well fermented), standard (S, partially fermented), and ordinary (O, underfermented). The fermentation ratio was calculated as $V_{\text{fer}} = (P + S)/ N$, where $N$ is the total number of beans, $P, S \ge 0$, and $P + S \le N$. This allowed labeling of each batch with a ground-truth fermentation level.}

\kebin{\textbf{Stage 6 – Batch Division:} Posteriorly, each 1.5~kg batch was divided into two portions: 600~g were set aside for \bacca{chemical} analysis, while 900~g were allocated for spectral signature acquisition. The spectral portion was further split into training and testing subsets (70\% and 30\%, respectively), ensuring physical separation to avoid data leakage.}

\kebin{\textbf{Stage 7 – Chemical Analysis:} The 600~g subsample reserved for chemical analysis was processed at the \textit{“Ciencia y Tecnología de Alimentos”} \citep{CICTA_UIS_GrupLAC_2025} research group from the Universidad Industrial de Santander (UIS). Moisture content was measured using a gravimetric oven-drying method, cadmium concentration was determined via microwave-assisted atomic absorption spectroscopy, and polyphenol content was quantified using the Folin-Ciocalteu colorimetric assay as explained in the methodology part.}


\subsection*{Spectral Optical System}

\begin{figure}[H]
    \centering
    \includegraphics[width=1\linewidth]{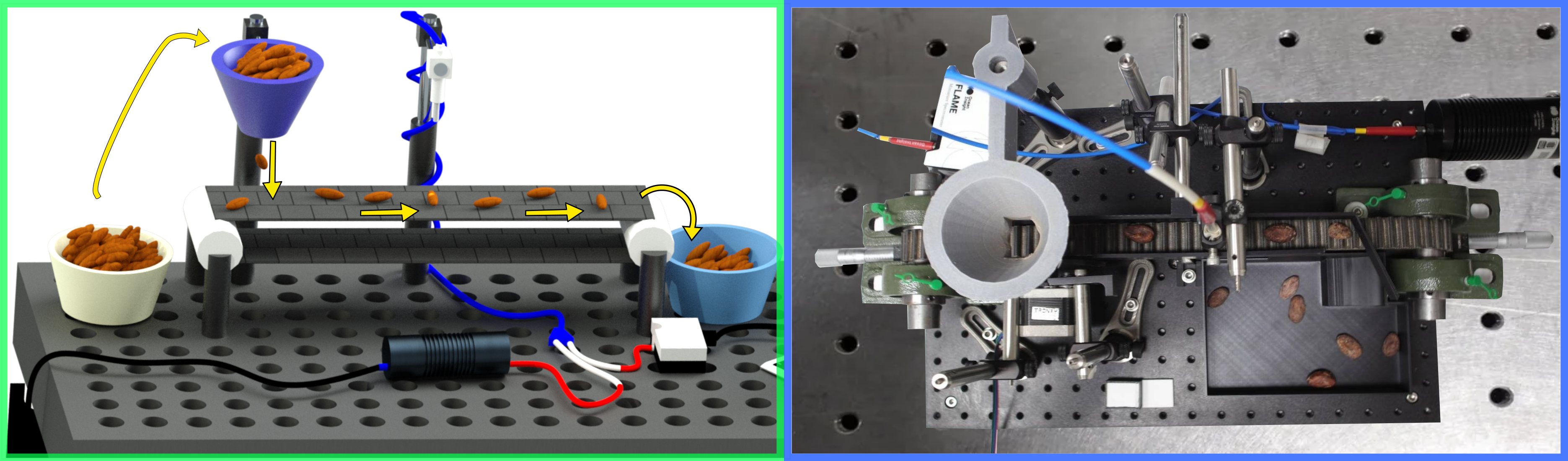}
    \caption{Automated system for spectral acquisition of cocoa beans. The left panel shows the system design, and the right panel shows its actual construction viewed from above: the conveyor belt, the light source, and the spectrometer for noninvasive data acquisition.}
    \label{fig:OpticalSystem}
\end{figure}

\kebin{A dedicated optical system was designed for the spectral acquisition of Colombian cocoa beans, operating at a throughput of up to 113 beans per minute under controlled lighting and acquisition parameters, as shown in Fig.~\ref{fig:OpticalSystem}. Each bean is positioned within a fixed measurement zone, where it is illuminated by a halogen lamp and scanned in both visible (VIS: 400–700 nm) and near-infrared (NIR: 700–2500 nm) spectral ranges.} 

\kebin{The system employs a bifurcated optical fibre that simultaneously delivers illumination and collects reflected light. The fibre is mounted at a fixed vertical distance of 14 cm from the conveyor belt. This configuration results in an average surface illumination of 59\%, with stable acquisition geometry across measurements. The spectrometer connection is modular, allowing rapid switching between VIS and NIR devices without requiring adjustments to the optical alignment or acquisition setup.}

\subsection*{Cocoa Bean Dataset}

\begin{table}[!t]
\begin{minipage}{\textwidth}
\captionof{table}{Summary of physicochemical properties of Colombian cocoa bean samples grouped by date of receipt. The table includes fermentation level, moisture content, cadmium concentration, polyphenol content, and fermentation time for each lot. Grouping by date highlights variations across reception periods, potentially reflecting differences in post-harvest handling, origin, or environmental conditions.}
\label{tab:grouped_labels_colormap}
\footnotesize
\centering

\begin{tabular}{
>{\centering\arraybackslash}p{1em}
>{\centering\arraybackslash}p{7em}
>{\centering\arraybackslash}p{7em}
>{\centering\arraybackslash}p{7em}
>{\centering\arraybackslash}p{7em}
>{\centering\arraybackslash}p{7em}
}
\toprule
\makecell{\textbf{\#}} &
\makecell{\textbf{Fermentation} \\ \textbf{Level [\%]}} &
\makecell{\textbf{Moisture} \\ \textbf{[\%]}} &
\makecell{\textbf{Cadmium} \\ \textbf{[mg/kg]}} &
\makecell{\textbf{Polyphenols} \\ \textbf{[mg/g]}} &
\makecell{\textbf{Fermentation} \\ \textbf{Time [h]}} \\
\midrule

\multicolumn{6}{l}{\textbf{Date of Receipt: 15/04/2024}} \\
1 & \colorFermentation{60} & \colorMoisture{5.12} & \colorCadmium{2.14} & \colorPolyphenols{41.30} & \colorHours{96} \\
2 & \colorFermentation{66} & \colorMoisture{4.93} & \colorCadmium{1.29} & \colorPolyphenols{34.24} & \colorHours{144} \\
3 & \colorFermentation{84} & \colorMoisture{4.80} & \colorCadmium{1.25} & \colorPolyphenols{40.38} & \colorHours{264} \\
4 & \colorFermentation{92} & \colorMoisture{4.75} & \colorCadmium{1.23} & \colorPolyphenols{39.81} & \colorHours{264} \\

\multicolumn{6}{l}{\textbf{Date of Receipt: 27/06/2024}} \\
5 & \colorFermentation{73} & \colorMoisture{4.79} & \colorCadmium{2.57} & \colorPolyphenols{32.85} & \colorHours{144} \\
6 & \colorFermentation{85} & \colorMoisture{4.94} & \colorCadmium{1.69} & \colorPolyphenols{39.75} & \colorHours{110} \\
7 & \colorFermentation{94} & \colorMoisture{4.56} & \colorCadmium{2.19} & \colorPolyphenols{28.78} & \colorHours{216} \\
8 & \colorFermentation{96} & \colorMoisture{5.09} & \colorCadmium{1.73} & \colorPolyphenols{23.74} & \colorHours{252} \\

\multicolumn{6}{l}{\textbf{Date of Receipt: 22/10/2024}} \\
9 & \colorFermentation{66} & \colorMoisture{5.82} & \colorCadmium{5.55} & \colorPolyphenols{27.66} & \colorHours{96} \\
10 & \colorFermentation{94} & \colorMoisture{5.68} & \colorCadmium{4.80} & \colorPolyphenols{23.05} & \colorHours{144} \\
11 & \colorFermentation{96} & \colorMoisture{5.67} & \colorCadmium{3.65} & \colorPolyphenols{25.09} & \colorHours{216} \\
12 & \colorFermentation{100} & \colorMoisture{5.67} & \colorCadmium{3.14} & \colorPolyphenols{22.76} & \colorHours{252} \\

\multicolumn{6}{l}{\textbf{Date of Receipt: 22/11/2024}} \\
13 & \colorFermentation{30} & \colorMoisture{6.68} & $<$\colorCadmium{0.09} & \colorPolyphenols{35.41} & \colorHours{30} \\
14 & \colorFermentation{45} & \colorMoisture{6.60} & $<$\colorCadmium{0.09} & \colorPolyphenols{37.29} & \colorHours{45} \\
15 & \colorFermentation{70} & \colorMoisture{6.87} & $<$\colorCadmium{0.09} & \colorPolyphenols{36.48} & \colorHours{70} \\
16 & \colorFermentation{70} & \colorMoisture{8.44} & $<$\colorCadmium{0.09} & \colorPolyphenols{25.90} & \colorHours{70} \\


\midrule

\multicolumn{6}{l}{\textbf{Date of Receipt: 18/01/2025}} \\
17 & \colorFermentation{44} & \colorMoisture{4.78} & \colorCadmium{2.65} & \colorPolyphenols{39.16} & \colorHours{30} \\
18 & \colorFermentation{70} & \colorMoisture{4.88} & \colorCadmium{2.72} & \colorPolyphenols{16.69} & \colorHours{45} \\
19 & \colorFermentation{87} & \colorMoisture{5.01} & \colorCadmium{2.24} & \colorPolyphenols{35.77} & \colorHours{70} \\
20 & \colorFermentation{96} & \colorMoisture{4.16} & \colorCadmium{1.7} & \colorPolyphenols{37} & \colorHours{70} \\

\bottomrule
\end{tabular}

\vspace{1em}

\begin{center}

\begin{minipage}{0.3\textwidth}
\centering
\begin{tikzpicture}
\begin{axis}[width=\textwidth, hide axis,
  colormap={fermentationmap}{rgb255=(252,244,240) rgb255=(200,150,100)},
  point meta min=30, point meta max=100,
  colorbar horizontal,
  colorbar style={
    height=0.3cm,
    xtick={40,60,80,100},
    xlabel=Fermentation Level [\%],
    tick label style={font=\small},
    xlabel style={font=\small},
    xtick style={draw=none}
  }]
\addplot [draw=none] coordinates {(0,0)};
\end{axis}
\end{tikzpicture}
\end{minipage}
\begin{minipage}{0.3\textwidth}
\centering
\begin{tikzpicture}
\begin{axis}[width=\textwidth, hide axis,
  colormap={moisturemap}{rgb255=(255,255,240) rgb255=(255,220,90)},
  point meta min=4, point meta max=9,
  colorbar horizontal,
  colorbar style={
    height=0.3cm,
    xtick={4,5,6,7,8,9},
    xlabel=Moisture [\%],
    tick label style={font=\small},
    xlabel style={font=\small},
    xtick style={draw=none}
  }]
\addplot [draw=none] coordinates {(0,0)};
\end{axis}
\end{tikzpicture}
\end{minipage}
\begin{minipage}{0.3\textwidth}
\centering
\begin{tikzpicture}
\begin{axis}[width=\textwidth, hide axis,
  colormap={cadmiummap}{rgb255=(255,252,252) rgb255=(245,120,120)},
  point meta min=0, point meta max=6,
  colorbar horizontal,
  colorbar style={
    height=0.3cm,
    xtick={0,1,2,3,4,5,6},
    xlabel=Cadmium [mg/kg],
    tick label style={font=\small},
    xlabel style={font=\small},
    xtick style={draw=none}
  }]
\addplot [draw=none] coordinates {(0,0)};
\end{axis}
\end{tikzpicture}
\end{minipage}

\vspace{0.5em}

\begin{minipage}{0.4\textwidth}
\centering
\begin{tikzpicture}
\begin{axis}[width=\textwidth, hide axis,
  colormap={polyphenolmap}{rgb255=(245,255,245) rgb255=(100,200,100)},
  point meta min=20, point meta max=45,
  colorbar horizontal,
  colorbar style={
    height=0.3cm,
    xtick={20,25,30,35,40,45},
    xlabel=Polyphenols [mg/g],
    tick label style={font=\small},
    xlabel style={font=\small},
    xtick style={draw=none}
  }]
\addplot [draw=none] coordinates {(0,0)};
\end{axis}
\end{tikzpicture}
\end{minipage}
\hspace{2em}
\begin{minipage}{0.4\textwidth}
\centering
\begin{tikzpicture}
\begin{axis}[width=\textwidth, hide axis,
  colormap={hourmap}{rgb255=(245,255,255) rgb255=(110,200,200)},
  point meta min=0, point meta max=300,
  colorbar horizontal,
  colorbar style={
    height=0.3cm,
    xtick={0,50,100,150,200,250,300},
    xlabel=Fermentation Time [h],
    tick label style={font=\small},
    xlabel style={font=\small},
    xtick style={draw=none}
  }]
\addplot [draw=none] coordinates {(0,0)};
\end{axis}
\end{tikzpicture}
\end{minipage}

\end{center}
\end{minipage}
\end{table}

\begin{table}[!t]
\begin{minipage}{\textwidth}
\captionof{table}{Summary of physicochemical properties of cocoa bean samples grouped by regions. The table includes fermentation level, moisture content, cadmium concentration, and polyphenol content for each lot. Grouping by region highlights variations across Latin America zones, potentially reflecting differences in post-harvest handling, origin, or environmental conditions.}
\label{tab:region_dataset}
\footnotesize
\centering

\begin{tabular}{
>{\centering\arraybackslash}p{4em}
>{\centering\arraybackslash}p{4em}
>{\centering\arraybackslash}p{4em}
>{\centering\arraybackslash}p{6em}
>{\centering\arraybackslash}p{6em}
>{\centering\arraybackslash}p{6em}
>{\centering\arraybackslash}p{6em}
}
\toprule
\makecell{\textbf{Date of} \\ \textbf{Receipt}} & \makecell{\textbf{Region}} & \makecell{\textbf{Country}} &
\makecell{\textbf{Fermentation} \\ \textbf{Level [\%]}} &
\makecell{\textbf{Moisture} \\ \textbf{[\%]}} &
\makecell{\textbf{Cadmium} \\ \textbf{[mg/kg]}} &
\makecell{\textbf{Polyphenols} \\ \textbf{[mg/g]}} \\
\midrule

01/06/2025 & Santander & Colombia & \colorFermentation{96} & \colorMoisture{4.16} & \colorCadmium{1.7} & \colorPolyphenols{37} \\

06/06/2025 & Huila & Colombia & \colorFermentation{60} & \colorMoisture{5.02} & \colorCadmium{0.84} & \colorPolyphenols{35.21} \\

10/06/2025 & Putumayo & Colombia & \colorFermentation{96} & \colorMoisture{5.02} & $<$\colorCadmium{0.09} & \colorPolyphenols{28.80} \\

12/06/2025 & Cusco & Peru & \colorFermentation{100} & \colorMoisture{5.36} & \colorCadmium{2.52} & \colorPolyphenols{33.94} \\

\bottomrule
\end{tabular}


\end{minipage}
\end{table}

To build the cocoa dataset, we processed 20 cocoa bean batches following the presented protocol. These batches were collected over nine months,
where four batches were gathered per sampling date. A detailed summary of the processed batches' physicochemical properties is provided in Table~\ref{tab:grouped_labels_colormap}. The table presents the fermentation level, moisture content, cadmium concentration, polyphenol content, and fermentation time for each batch, organized by the date of receipt. This grouping highlights potential variations across different reception periods, which may reflect differences in post-harvest handling, environmental conditions, or the origin of the beans. The cocoa batches with the date of receipt 18/01/2025 are used only for evaluation purposes, while other batches are used for training regression-based models.

Following the same experimental protocol, three additional cocoa bean batches were collected from Huila and Putumayo in southern Colombia and from the Cusco region in Peru, as shown in Tab. \ref{tab:region_dataset}. These additional batches were introduced in the test dataset to assess cocoa bean quality under diverse harvest and post-harvest conditions, such as differences in climate, fermentation practices, and drying methods. This variability is expected to enrich the analysis and improve the generalizability of the learning-based spectral regression models for cocoa bean quality assessment.

\begin{figure}[t]
    \centering
    \includegraphics[width=1\linewidth]{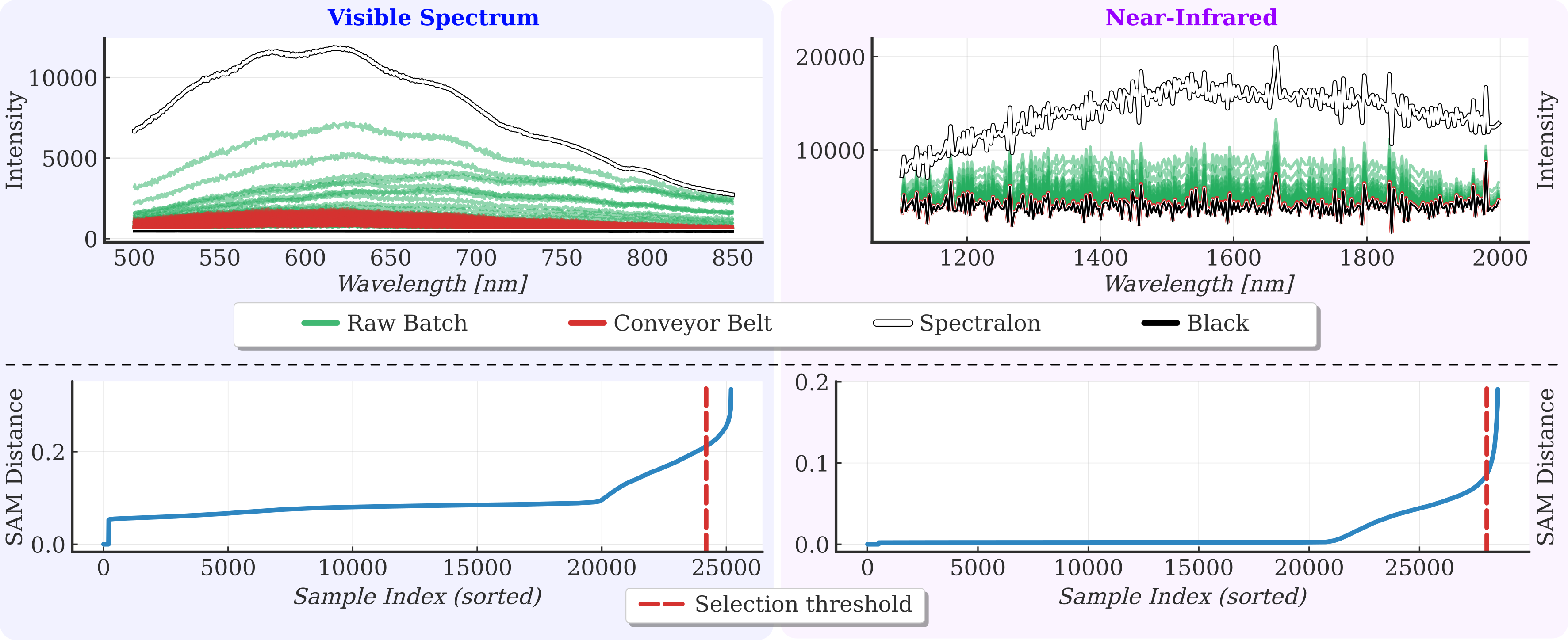}
    \caption{\textbf{Raw spectral dataset.} Visible (left) and near-infrared (right) intensity spectra for cocoa (green), background (red), and reference targets (black, white). Bottom: SAM distances to conveyor references; red dashed line indicates selection threshold.}
    \label{fig:dataset_raw}
\end{figure}

After acquisition with the proposed optical system, the spectral signatures were analyzed to identify the wavelength ranges of interest and remove regions affected by noise extremes. The VIS spectrum was defined from 500–850 nm, and the NIR spectrum from 1100–2000 nm. To isolate cocoa bean spectral signatures from background signals introduced by the conveyor belt, as shown in Fig.~\ref{fig:dataset_raw}, a Spectral Angle Mapper (SAM)~\citep{kruse1993spectral} distance metric was computed between known conveyor belt spectra and all raw samples in the intensity domain. This allowed quantitative discrimination between sample and background spectra. A threshold-based selection strategy was applied to retain only the most distinct spectral signatures. Specifically, the $n$ samples with the highest SAM distances were selected: $n = 1000$ for VIS and $n = 500$ for NIR. This filtering step aimed to exclude the most representative conveyor belt contributions and retain the cocoa-specific signatures for further analysis.

\begin{figure}[!t]
    \centering
    \includegraphics[width=0.95\linewidth]{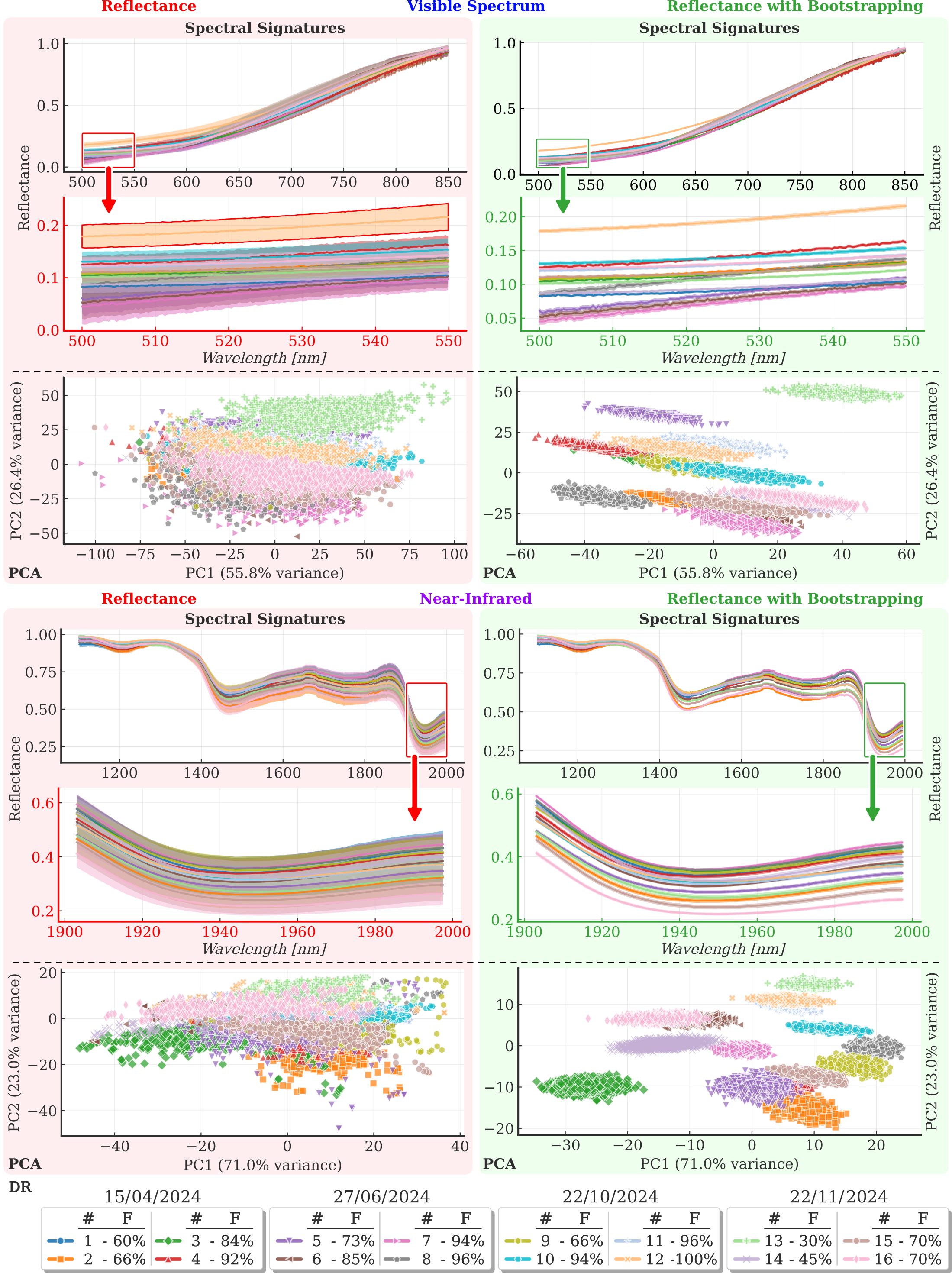}
    \caption{\textbf{Raw reflectance vs. reflectance with bootstrapping.} Top: reflectance spectra; bottom: PCA projections of cocoa samples from four 2024 dates. Raw data (left) show overlap, especially at 500–550 nm. Bootstrapped averaging (right) reveals clearer batch clustering. DR: Date of Receipt.}
    \label{fig:dataset_processed}
\end{figure}

Spectral variability within the selected cocoa samples was assessed across acquisition batches. The principal Component Analysis (PCA) revealed substantial spectral overlap between batches, as shown at the left of Fig.~\ref{fig:dataset_processed}, indicating a high degree of correlation and limited spectral separation. Such redundancy is undesirable for robust downstream modeling of physicochemical properties. Hence, to reduce variance and enhance discriminability, a bootstrapping \citep{zoubir2007bootstrap} approach was adopted, consistent with Colombian standard NTC 1252:2021. For each acquisition batch, 50 spectral signatures were randomly selected and averaged spatially. This process was repeated 1000 times for VIS and 2000 times for NIR, generating representative mean profiles while attenuating intra-batch variability. The resulting datasets exhibited markedly improved spectral differentiation across batches, particularly in the 500–550 nm and 1950–2000 nm ranges for VIS and NIR, respectively, as shown at the right of Fig.~\ref{fig:dataset_processed}. Corresponding PCA projections with 2 components confirmed enhanced clustering and separation, facilitating more robust interpretation and subsequent modeling.

\subsection*{Regression-based Cocoa Bean Quality Assessment}

A variety of machine learning and deep learning models \citep{smola2004tutorial, breiman2001random, hochreiter1997long, vaswani2017attention} were assessed to predict four physicochemical attributes of cocoa beans: cadmium concentration, moisture content, fermentation level, and polyphenol content. The models were trained and evaluated using spectral data from the VIS and NIR regions. Table~\ref{tab:r2_mse_comparison} presents a summary of the results in terms of coefficient of determination ($\mathcal{R}^2$) and mean squared error (MSE).

Machine learning models such as SVR, RFR, and KNNR produced lower and more variable predictive results. SVR and RFR yielded $\mathcal{R}^2$ values below 0.7 for most targets. KNNR showed better performance for cadmium and moisture using NIR data, with $\mathcal{R}^2$ values of 0.9081 and 0.9740, respectively. However, their results were less consistent than those obtained with deep learning models.

\begin{table}[h]
\caption{\textbf{Comparison of $\mathcal{R}^2$ and $MSE$ for Deep Learning and Machine Learning Models using the designed dataset.} The best and second-best performances per column are highlighted in green (bold) and blue (underline), respectively, per model category (Deep Learning and Machine Learning), regardless of spectral range.}
\label{tab:r2_mse_comparison}
\footnotesize
\centering
\renewcommand{\arraystretch}{1.2}

\begin{tabular}{
>{\centering\arraybackslash}p{3em}
>{\centering\arraybackslash}p{2.2em}
*{4}{>{\centering\arraybackslash}p{3.5em}}  
*{4}{>{\centering\arraybackslash}p{3.5em}}  
}
\toprule
\textbf{Method} & \textbf{Range} &
\multicolumn{4}{c}{\textbf{$\mathcal{R}^2$}} &
\multicolumn{4}{c}{$MSE$} \\
\cmidrule(lr){3-6} \cmidrule(lr){7-10}
& & \textbf{Cadmium} & \textbf{Moisture} & \textbf{Ferment.} & \textbf{Polyph.} 
  & \textbf{Cadmium} & \textbf{Moisture} & \textbf{Ferment.} & \textbf{Polyph.} \\
\midrule
\multicolumn{10}{l}{\textbf{Deep Learning}} \\
S-Net & VIS & \cellcolor{green!15}\textbf{0.9605} & 0.9665 & 0.8457 & 0.8684 & \cellcolor{green!15}\textbf{0.045} & 0.032 & 0.054 & 0.043 \\
& NIR & 0.9449 & 0.9522 & \cellcolor{blue!15}\underline{0.8718} & 0.8619 & 0.050 & 0.034 & \cellcolor{blue!15}\underline{0.052} & 0.041 \\
CNN  & VIS & 0.8308 & 0.4032 & 0.8128 & 0.5289 & 0.134 & 0.128 & 0.097 & 0.115 \\
& NIR & 0.8232 & 0.7697 & 0.7863 & 0.7180 & 0.140 & 0.100 & 0.095 & 0.098 \\
LSTM & VIS & 0.9070 & 0.9555 & \cellcolor{green!15}\textbf{0.8718} & 0.8205 & 0.088 & 0.038 & \cellcolor{green!15}\textbf{0.050} & 0.045 \\
& NIR & 0.8696 & 0.9681 & 0.8429 & 0.8457 & 0.098 & 0.036 & 0.055 & 0.047 \\
Transf. & VIS & 0.9558 & \cellcolor{blue!15}\underline{0.9824} & 0.7965 & \cellcolor{blue!15}\underline{0.8900} & 0.048 & \cellcolor{blue!15}\underline{0.025} & 0.062 & \cellcolor{blue!15}\underline{0.039} \\
& NIR & \cellcolor{blue!15}\underline{0.9590} & \cellcolor{green!15}\textbf{0.9926} & 0.7671 & \cellcolor{green!15}\textbf{0.9006} & \cellcolor{blue!15}\underline{0.046} & \cellcolor{green!15}\textbf{0.022} & 0.065 & \cellcolor{green!15}\textbf{0.037} \\
\midrule
\multicolumn{10}{l}{\textbf{Machine Learning}} \\
SVR & VIS & 0.6321 & 0.3034 & 0.6168 & 0.3214 & 0.250 & 0.270 & 0.190 & 0.230 \\
& NIR & 0.6729 & 0.3167 & 0.6058 & 0.3407 & 0.240 & 0.265 & 0.185 & 0.225 \\
RFR & VIS & 0.6020 & 0.8592 & 0.6670 & 0.7992 & 0.270 & 0.070 & 0.140 & 0.120 \\
& NIR & 0.6175 & 0.8701 & 0.6707 & \cellcolor{green!15}\textbf{0.8005} & 0.265 & 0.068 & 0.138 & 0.118 \\
KNNR & VIS & \cellcolor{blue!15}\underline{0.8901} & \cellcolor{blue!15}\underline{0.9365} & \cellcolor{blue!15}\underline{0.7232} & 0.7689 & \cellcolor{blue!15}\underline{0.090} & \cellcolor{blue!15}\underline{0.050} & \cellcolor{blue!15}\underline{0.110} & \cellcolor{blue!15}\underline{0.098} \\
& NIR & \cellcolor{green!15}\textbf{0.9081} & \cellcolor{green!15}\textbf{0.9740} & \cellcolor{green!15}\textbf{0.7372} & \cellcolor{blue!15}\underline{0.8002} & \cellcolor{green!15}\textbf{0.085} & \cellcolor{green!15}\textbf{0.045} & \cellcolor{green!15}\textbf{0.108} & \cellcolor{green!15}\textbf{0.095} \\
\bottomrule
\end{tabular}
\end{table}

\subsubsection*{Generalization on Geographically Independent Samples}

To evaluate model generalization, performance was tested on external cocoa batches sourced from various regions as shown in Table \ref{tab:region_dataset}. Reference values were obtained from standardized laboratory analyses. As shown in Table~\ref{tab:industrial_test}, predictions were computed using the best model of ML and DL models with both VIS and NIR spectra. In the international batch from Peru, the most accurate polyphenol prediction was obtained with VIS-DL (29.43 mg/g, reference: 28.80 mg/g). Cadmium was best estimated using VIS-DL (0.08 mg/kg, reference: 0.09 mg/kg). Fermentation was also best predicted by VIS-DL (93.64\%, reference: 96.00\%). For moisture, NIR-DL yielded the closest result (4.97\%, reference: 5.02\%). 

\begin{figure}[hbt]
    \centering
    \includegraphics[width=1\linewidth]{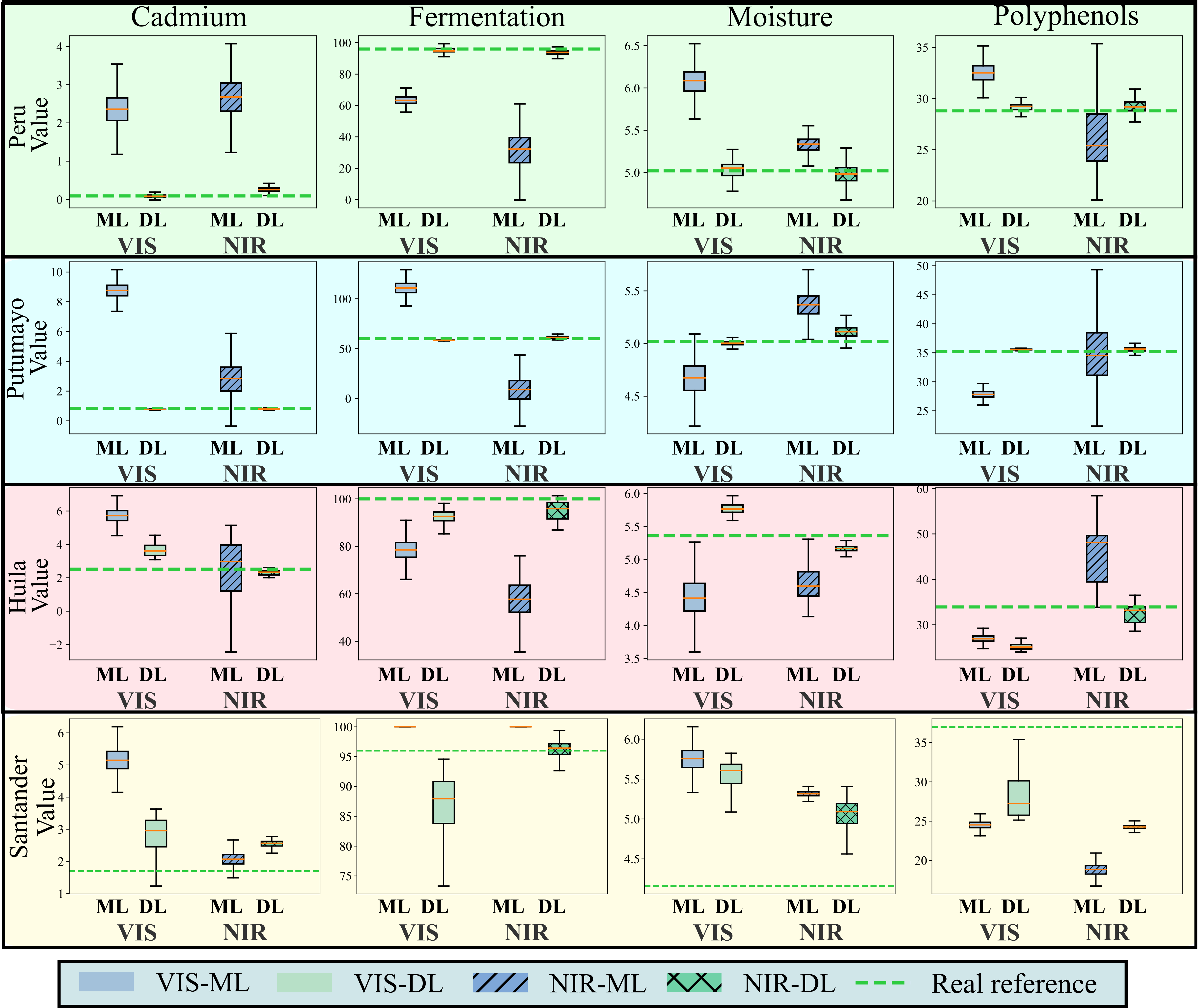}
    \caption{Predicted versus laboratory values for external cocoa batches using ML and DL models with VIS and NIR spectra. The green dashed line indicates the reference value.}

    \label{fig:fig7_global_results}
\end{figure}

Fig.~\ref{fig:fig7_global_results} illustrates the prediction results for each region and configuration. These findings indicate that the models trained with data from Santander were able to produce accurate estimations when applied to batches from other origins, under varying post-harvest conditions. This result suggests that the spectral and compositional variability of Santander cocoa encompasses representative patterns that are also present in beans from other regions, enabling the models to generalize effectively. In addition, the consistency of predictions across different physicochemical properties demonstrate the robustness of the proposed approach. Although some fluctuations appear depending on the attribute and the spectral domain (VIS or NIR), the overall trend confirms that both machine learning and deep learning models capture key features that remain stable across origins.

\begin{table}
\caption{\textbf{Generalization on different region cocoa batches.} The closest prediction to the real lab value is highlighted in green, the second closest in blue.}
\label{tab:industrial_test}
\footnotesize
\renewcommand{\arraystretch}{1.2}
\centering
\begin{tabular}{
>{\centering\arraybackslash}p{4em}
>{\centering\arraybackslash}p{7em}
>{\centering\arraybackslash}p{4em}
>{\centering\arraybackslash}p{4em}
>{\centering\arraybackslash}p{4em}
>{\centering\arraybackslash}p{4em}
>{\centering\arraybackslash}p{4em}
}
\toprule
\multicolumn{3}{c}{} & \multicolumn{4}{c}{\textbf{Predicted of our model}} \\
\cmidrule(lr){4-7}
\textbf{Batch} & \textbf{Property} & \textbf{Chemical Labels} & \textbf{VIS ML} & \textbf{NIR ML} & \textbf{VIS DL} & \textbf{NIR DL} \\
\midrule
\multirow{4}{*}{\rotatebox[origin=c]{90}{Peru}} 
  & Cadmium & 0.09 & 2.36 & 2.68 & \cellcolor{green!15}\textbf{0.08} & \cellcolor{blue!15}\underline{0.26} \\
  & Fermentation    & 96.00 & 63.46 & 31.06 & \cellcolor{green!15}\textbf{93.64} & \cellcolor{blue!15}\underline{93.43} \\
  & Moisture & 5.02 & 6.08 & 5.33 & \cellcolor{blue!15}\underline{5.07} & \cellcolor{green!15}\textbf{4.97} \\
  & Polyphenols & 28.80 & 32.55 & 26.39 & \cellcolor{blue!15}\underline{29.43} & \cellcolor{green!15}\textbf{29.28} \\
\midrule
\multirow{4}{*}{\rotatebox[origin=c]{90}{Putumayo}} 
  & Cadmium & 0.84 & 8.76 & 2.76 & \cellcolor{blue!15}\underline{0.77} & \cellcolor{green!15}\textbf{0.80} \\
  & Fermentation    & 60.00 & 99.91 & 8.68 & \cellcolor{green!15}\textbf{58.44} & \cellcolor{blue!15}\underline{61.67} \\
  & Moisture & 5.02 & 4.67 & 5.36 & \cellcolor{green!15}\textbf{5.00} & \cellcolor{blue!15}\underline{5.11} \\
  & Polyphenols & 35.21 & 27.89 & \cellcolor{green!15}\textbf{35.02} & 35.58 & \cellcolor{blue!15}\underline{35.54} \\
\midrule
\multirow{4}{*}{\rotatebox[origin=c]{90}{Huila}} 
  & Cadmium & 2.52 & 5.73 & \cellcolor{green!15}\textbf{2.59} & 3.66 & \cellcolor{blue!15}\underline{2.30} \\
  & Fermentation    & 100.00 & 78.53 & 57.17 & \cellcolor{blue!15}\underline{92.49} & \cellcolor{green!15}\textbf{95.19} \\
  & Moisture & 5.36 & 4.44 & 4.62 & \cellcolor{blue!15}\underline{5.77} & \cellcolor{green!15}\textbf{5.16} \\
  & Polyphenols & 33.94 & \cellcolor{blue!15}\underline{27.01} & 45.66 & 25.22 & \cellcolor{green!15}\textbf{32.55} \\
\midrule
\multirow{4}{*}{\rotatebox[origin=c]{90}{Santander}} 
  & Cadmium & 1.70 & \cellcolor{blue!15}\underline{2.07} & 2.79 & 2.54 & \cellcolor{green!15}\textbf{1.70} \\
  & Fermentation    & 96.00 & \cellcolor{blue!15}\underline{99.98} & 86.59 & \cellcolor{green!15}\textbf{96.01} & 100.00 \\
  & Moisture & 4.16 & 5.75 & 5.31 & 5.54 & \cellcolor{green!15}\textbf{5.07} \\
  & Polyphenols & 37.00 & \cellcolor{green!15}\textbf{28.17} & 24.33 & 24.52 & \cellcolor{blue!15}\underline{28.07} \\
\bottomrule
\end{tabular}
\end{table}

\section*{Conclusion}\label{Conclusion}

This study presents a scalable and non-destructive approach for assessing cocoa bean quality using VIS and NIR spectroscopy with learning-based regression models. The proposed models demonstrated predictive performance for key physicochemical properties, outperforming traditional ML techniques. These findings underscore the potential of DL architectures, particularly Transformer and SpectralNet, for modeling spectral data in agricultural applications. \textcolor{black}{Beyond overall outperformance, the study identifies property–range leaders: Transformer achieves accuracy for moisture and polyphenols using NIR spectra; SpectralNet provides the best cadmium estimation using VIS spectra; and LSTM attains the strongest fermentation prediction in VIS, enabling a property-specific deployment strategy.} \textcolor{black}{Moreover, the proposed acquisition and spectral curation pipeline, combining background discrimination and intra-batch bootstrapped averaging, improves inter-batch separability and stabilizes regressors, constituting an enabler for in-line quality control.}

The integration of VIS–NIR spectroscopy with learning-based regression enables rapid, non-invasive evaluation of cocoa quality. This approach provides an alternative to destructive laboratory analyses, reducing time, cost, and resource demands while maintaining compliance with international quality standards. Generalization tests conducted on cocoa samples from regions such as Huila and Putumayo in Colombia, and Cusco in Peru, confirmed that models trained on Santander data retain performance across varied geographic, environmental, and post-harvest conditions. These results validate the transferability and reliability of the models in real-world scenarios. \textcolor{black}{Crucially, the cross-regional evaluation evidences geographic transfer with errors competitive relative to local laboratory references, indicating that a model trained in one producing area can be ported to distinct agroecological contexts with minimal recalibration.} This dataset will be made available to support future research aimed at improving the prediction of cocoa bean physicochemical properties from spectral information. \textcolor{black}{Collectively, these findings position VIS–NIR plus DL as a viable, scalable substitute for destructive assays in post-harvest quality control, enabling faster decision cycles and broader coverage at lower operational cost.}

Despite these outcomes, certain challenges must be addressed to enable large-scale industrial adoption. Future work should focus on integrating spectral acquisition systems into on-site processing environments and developing hybrid models that combine spectral and image-based features. These enhancements could further improve model accuracy and adaptability, supporting the implementation of sustainable, data-driven quality control systems within the cocoa industry. \textcolor{black}{Further research should also quantify calibration-transfer requirements across devices and sites, and assess active learning schemes to maintain accuracy under drift in post-harvest practices and seasonal shifts.}

\section*{Acknowledgements}\label{Acknowledgements}

Additionally, the authors thank Miguel Belmin for contributing to the standardization of the cocoa acquisition protocol and for assisting in the labeling of fermentation properties, Ana Celina Gutiérrez for administrative assistance throughout the project. Appreciation is also extended to Laura Camila Díaz-Delgado for support in the acquisition of VIS–NIR spectral data.

\section*{Declarations}


\begin{itemize}
\item Funding:  The authors acknowledge the support of the Vicerrectoría de Investigación y Extensión of Universidad Industrial de Santander (VIE-UIS) for funding this work through the project 3924.

\item Conflict of interest/Competing interests: The authors declare no competing interests. 





\item Code and Data availability: The code and associated dataset are publicly available at \url{https://github.com/PIgroupUIS/SpecCocoa_Regression_Physicochemical_Properties.git}

\item Author contribution: Author initials used in the table above correspond to the following full names: KC = Kebin Contreras, EM = Emmanuel Martinez, BM = Brayan Monroy, SA = Sebastian Ardila, CR = Cristian Ramirez, MC = Mariana Caicedo, HG = Hans Garcia, TG = Tatiana Gelvez-Barrera, JPJ = Juan Poveda-Jaramillo, HA = Henry Arguello, JB = Jorge Bacca.

\begin{tabular}{
>{\centering\arraybackslash}p{5cm}
>{\centering\arraybackslash}p{7cm}
}
\toprule \toprule
    \textbf{Contribution} & \textbf{Authors} \\
\midrule

Conceptualization & KC, EM, BM, SA, CR, MC, HG, TG, JPJ, HA, JB \\ \hline
Methodology & KC, EM, SA, MB, JB \\ \hline
Data acquisition & KC, EM, BM, HG, SA \\ \hline
Software & KC, EM \\ \hline
Visualization & KC, EM, JB \\ \hline
Data Curation & KC, EM, BM \\ \hline
Writing -- Original Draft & KC, EM, SA \\ \hline
Writing -- Review \& Editing & BM, TG, JB \\ \hline
Regulatory Standards Analysis & CR, MC \\ \hline
Supervision / General Review & HG, JPJ, JB \\
\bottomrule \bottomrule \\
\end{tabular}

\end{itemize}

\noindent







\bibliography{sn-article.bib}

\end{document}